# FROGi : Déploiement de composants Fractal sur OSGi

## Humberto Cervantes, Mikael Désertot, Didier Donsez

*Universidad Autonoma Metropolitana-Iztapalapa,*
*San Rafael Atlixco N° 186, Col. Vicentina, C.P. 09340*
*Delegación Iztapalapa. Distrito Federal, Mexico*

*Fédération IMAG, Laboratoire LSR, Equipe ADELE*
*220 rue de la Chimie, Domaine Universitaire, BP 53*
*38041 Grenoble, Cedex 9, FRANCE*
*{Humberto.Cervantes, Mikael.Desertot, Didier.Donsez}@imag.fr*

*RÉSUMÉ. Cet article présente FROGi, une proposition visant à introduire le modèle à composants Fractal à l'intérieur de la plateforme de services OSGi. La motivation derrière ce travail est double. D'un côté, FROGi offre aux développeurs de services OSGi un modèle à composants extensibles qui facilite le développement des bundles ; ces derniers restent toutefois compatibles avec les bundles « patrimoniaux ». D'un autre côté, FROGi bénéficie de l'infrastructure de déploiement que représente OSGi et qui facilite la réalisation du conditionnement et du déploiement de composants Fractal. Dans FROGi, une application Fractal est conditionnée sous la forme d'un ou plusieurs bundles et elle peut être déployée de façon partielle et les activités de déploiement peuvent avoir lieu de façon continue.*

*MOTS-CLÉS : Développement à base de Composants, Déploiement, Conception orientée service, Fractal, OSGi.*

*ABSTRACT : This paper presents FROGi, a proposal to introduce the Fractal component model into the OSGi services platform. There are two motivations for this work. The first one is to offer a flexible component model to the OSGi developers to simplify bundle development. Bundles developed with FROGi are nevertheless compatible with standard bundles. The second motivation is to leverage OSGi's deployment capabilities to package and deploy Fractal components. In FROGi, a Fractal application is packaged and delivered as a set of OSGi bundles; such an application supports partial deployment and additionally, deployment activities can occur continuously.*

**1. Introduction**

L'approche à composants est une méthodologie qui soutient l'idée que la construction de logiciels peut être réalisée à partir de l'assemblage d'entités logicielles réutilisables appelées composants [SZY 98]. Un composant définit de manière explicite un ensemble de provisions et de dépendances pour permettre de réaliser sa composition. Un aspect important de l'approche à composants est qu'elle suppose que les activités de développement de composants et leur assemblage sont clairement différentiées, ces activités peuvent d'ailleurs être réalisées par des acteurs distincts. Cette différentiation nécessite que les aspects de livraison et de déploiement soient pris en compte tôt dans le cycle de vie de développement. Pour faciliter la livraison et le déploiement, un composant est typiquement conditionné dans un paquetage. Ce paquetage contient tout ce qui est nécessaire au composant pour fonctionner, à l'exception de ce que le composant déclare comme étant une dépendance explicite et qui est remplie soit à travers la composition soit au moment du déploiement. Actuellement il existe de nombreux modèles à composants. Parmi ces modèles, certains sont destinés à des domaines d'application spécifiques, tels que COM, EJB, CCM et .NET tandis que d'autres se veulent plus universels comme Fractal [BRU 04].

Cet article se concentre sur le modèle à composants Fractal et son implémentation de référence Julia [BRU 03]. En particulier, cet article propose FROGi, qui introduit dans Julia plusieurs éléments qui ne sont pas disponibles dans le modèle Fractal de base. Le premier de ces éléments est un moyen de réaliser le conditionnement et le déploiement de composants et de composites Fractal en utilisant des unités de déploiement appelées *bundles* définies par plateforme de services OSGi [OSG 03]. OSGi, qui est aujourd'hui le standard de fait des serveurs embarqués, à été conçu à l'origine pour faciliter le déploiement de services dans des passerelles résidentielles devant fonctionner sans interruption. Un deuxième élément introduit par FROGi est une infrastructure ainsi qu'un outillage étendu permettant de réaliser, de façon continue, la gestion du cycle de vie de déploiement, ce qui inclut des activités d'installation, activation, mise à jour et retrait des bundles [CAR 98]. Finalement, FROGi introduit aussi des mécanismes permettant de réaliser du courtage propre à l'approche orientée services [BIE 01].

FROGi combine des caractéristiques du modèle à composants Fractal ainsi que celles de la plateforme OSGi. Ce mélange permet non seulement de simplifier le déploiement d'applications Fractal, mais aussi de réaliser des applications dynamiques basées sur une approche orientée services. Cet article décrit les concepts et la réalisation de FROGi en essayant de montrer les difficultés rencontrées lors de sa réalisation.

Cet article est structuré de la façon suivante. La section 2 présente les bases de FROGi, qui sont Fractal et OSGi. La section 3 présente FROGi et décrit comment une application Fractal qui est conditionnée sous la forme de bundles, livrée et démarrée avec FROGi. La section 4 présente des travaux liés et finalement la section 5 présente une conclusion et des perspectives à ce travail.

**2. Bases**

Cette section présente le modèle à composants Fractal ainsi que la plate-forme de services OSGi.

*2.1. Fractal*

Fractal définit un framework (cadre de conception) permettant de construire des applications à partir d'un modèle à composants supportant la création de compositions hiérarchiques. Le framework Fractal fournit une API permettant, entre autres, de définir des types de composants ainsi que de créer des instances à partir de ces types, de les configurer et connecter entre elles. Un composant Fractal (voir figure 1) est caractérisé comme étant une entité qui fournit et requiert des interfaces fonctionnelles (c-a-d liées à la logique applicative représenté par les facettes jl.R, y.Y et z.Z) et qui fournit un ensemble d'interfaces de contrôle. La liste des interfaces de contrôle inclue, entre autres, une interface dédiée à la gestion du cycle de vie (*LifecycleController* -LC-), à la liaison entre instances (*BindingController* -BC-) et au contrôle du contenu (*ContentController* -CC-) dans le cas où le composant serait composite. Plusieurs instances d'un composant Fractal peuvent être créées à partir d'une fabrique associée à un type de composant. L'exemple de la figure 1 décrit un composant composite contenant deux instances de composants. Ces instances de composants, qui représentent un client et un serveur sont connectées et une interface fournie par l'instance client est exportée en dehors de la composition.

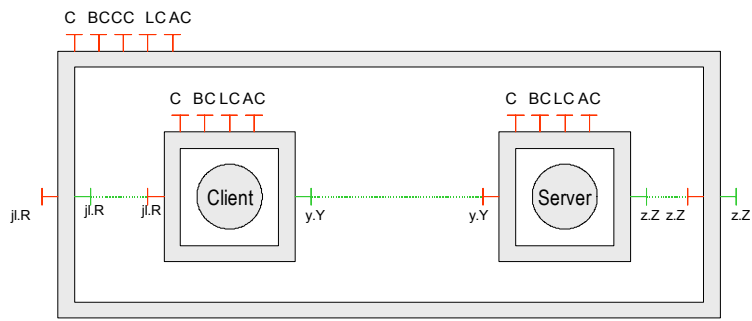

**Figure 1.** *Représentation graphique d'un composite.*

Julia [BRU 03] est l'implémentation de référence du framework Fractal pour des applications Java. Julia a pour but de simplifier la construction d'applications Fractal à travers la génération de classes de support. Ces classes sont nécessaires aux classes Java implémentant la logique applicative pour leur permettre d'adhérer au modèle de composants Fractal. Les classes de support sont générées de façon statique ou dynamique par Julia à travers des techniques de Mixin et d'injection de bytecode ;

elles incluent notamment des classes qui implémentent les interfaces de contrôle ainsi que des intercepteurs qui se placent entre les interfaces fonctionnelles et l'implémentation du composant.

Une application Fractal est typiquement conçue à partir d'un ensemble de classes qui implémentent la logique applicative contenue dans les composants, une ou plusieurs classes de coordination, ainsi qu'une classe primaire (bootstrap) chargée de réaliser le démarrage de l'application. Les classes de coordination ont pour responsabilité d'interagir avec le framework Fractal pour créer des types de composants, des instances et de les configurer et connecter pour assembler l'application. L'ADL fractal permet de simplifier l'écriture des classes de coordination en supportant la description des composants ainsi que l'architecture d'une application de façon déclarative.

## *2.2. OSGi*

La spécification OSGi [OSG 03] définit une plateforme Java qui supporte le déploiement d'unités de livraison, appelées *bundles*, à l'intérieur d'un environnement d'exécution. L'environnement d'exécution permet à des composants contenus dans les bundles d'interopérer suivant l'approche orientée services. Dans OSGi, chaque bundle est utilisé pour déployer un seul composant qui au moment de l'exécution donne lieu à une instance unique (singleton). L'environnement d'exécution fournit des mécanismes permettant de réaliser des activités continues de déploiement qui incluent l'installation, l'activation, la déactivation, la mise à jour et le retrait des bundles. De plus, l'environnement d'exécution prend en charge la gestion de dépendances de code qui doivent être satisfaites après l'installation d'un bundle pour permettre de réaliser son activation. L'activation d'un bundle déclenche la création de l'instance du composant déployé dans le bundle.

Physiquement, un bundle est un fichier JAR qui contient du code binaire ainsi que des ressources nécessaires au composant qui est déployé. Le bundle contient aussi un fichier manifeste et une classe d'activation. Le manifeste décrit, entre autres, des dépendances de code du composant, qui incluent l'import et export de packages Java. Les dépendances de code sont utilisées dans OSGi pour deux raisons qui sont le partage de librairies et le partage de code associé aux interfaces de service. Le code associé aux interfaces de service doit être partagé pour permettre aux composants d'interagir, ceci permet en plus de maintenir le code d'implémentation des composants privés.

L'environnement d'exécution de la plateforme OSGi fournit un registre de services qui permet aux instances de composants de publier des services ou de récupérer des services fournis par d'autres instances de composants. Dans OSGi, un service est publié à partir d'une interface de service, une référence vers le composant qui implémente le service et un ensemble de propriétés. Ces propriétés, de type clé-valeur, permettent aux clients de différentier deux offres de services ayant la même interface. De plus, le registre fournit des moyens de réaliser des recherches contraintes par un filtre en syntaxe LDAP basé sur ces propriétés. Du fait que l'enregistrement ou le retrait de services peut être réalisé à tout moment, le registre de services supporte l'envoi de notifications signalant des changements dans le registre. Elles sont nécessaires aux clients des services pour être avertis de l'arrivée ou du départ d'un service particulier. Dans OSGi, l'assemblage d'une application a lieu au moment de l'exécution comme résultat de l'interaction entre les composants et le registre de services.

## 3. FROGi

Cette section présente successivement les différents points de la proposition FROGi : les motivations, le conditionnement d'une application, la gestion à l'exécution et enfin le déclenchement de l'application sur la passerelle.

### *3.1. Motivations*

La motivation derrière FROGi est double. D'un côté, FROGi cherche à offrir aux développeurs de services OSGi un modèle à composants extensible pour faciliter le développement des bundles. Les bundles développés avec FROGi restent cependant compatibles avec les bundles « patrimoniaux » [1]: un composant FROGi peut implémenter des services utilisés par les bundles de la plateforme, et il peut aussi utiliser les services fournis par d'autres bundles (FROGi ou non). Par exemple, le serveur HTTP Comanche conçu avec Fractal pourrait être un bon candidat pour fournir le service *org.osgi.service.http.HttpService*. A son tour, le serveur HTTP Comanche doit être capable d'utiliser le service *org.osgi.service.log.LogService* pour journaliser les accès et les erreurs.

D'un autre côté, FROGi s'appuie sur l'infrastructure de déploiement dynamique que représente OSGi pour réaliser le conditionnement et de déploiement de composants Fractal. Il permet à une application Fractal de faire les mises à jour des versions d'interfaces de service et des implémentations d'une partie des composants FROGi sans arrêter l'application dans son ensemble. FROGi s'appuie sur les outils existants et qui permettent de réaliser l'administration, supervision et déploiement disponibles pour les différentes passerelles OSGi, tant commerciales (IBM SMF, ProSyst, …) que de sources ouvertes (Oscar, Knoplerfish).

### *3.2. Conditionnement*

Dans FROGi, une application Fractal est conditionnée sous la forme d'un ou plusieurs bundles. Il est important de souligner qu'à l'intérieur du même bundle, les composants FROGi sont liés entre eux à partir de l'approche Fractal standard, cependant, dans le cas où les composants sont livrés dans des bundles séparés, la liaison est réalisée à travers l'approche orientée services d'OSGi (ceci est décrit plus loin).

---

[1] Le terme de patrimonial englobe à la fois les bundles déjà développés pour OSGi ou *off the shelf* et les bundles développé sans Fractal / FROGi.

Du fait qu'une application Fractal est réalisée comme une composition hiérarchique, FROGi supporte le conditionnement indépendant des composants primitifs ainsi que des composites. Ceci permet de réaliser la livraison séparée ainsi que la réalisation des mises à jour indépendantes. L'exemple de la figure 2 présente le conditionnement effectué pour l'application décrite dans la figure 1. Dans l'exemple, chaque composant est conditionné dans un bundle différent : B0 pour le composite, B1 pour le client, B2 pour le serveur. Il faut noter cependant qu'un bundle peut "emballer" un ou plusieurs composants primitifs ou composites.

Les interfaces de service utilisées dans les liaisons entre les composants qui sont livrés dans des bundles séparés sont conditionnées dans un ou plusieurs bundles séparés (le bundle B3 dans l'exemple). Ceci permet de mettre à jour un bundle contenant une implémentation d'un service sans impact sur les autres bundles. En effet, lorsque les interfaces de service sont livrées avec l'implémentation, la mise à jour du bundle entraîne l'arrêt et le redémarrage (i.e. *refresh*) des bundles qui utilisent les interfaces de ce service. Ceci peut poser des problèmes dans des environnements à fonctionnement continu (c-a-d sans interruption de service). Il faut souligner que les interfaces de service représentent des contrats qui sont assez stables et n'évoluent que lentement alors que les implémentations peuvent évoluer de façon fréquente.

Dans FROGi, l'API Fractal ainsi que le runtime Julia sont conditionnés à l'intérieur d'un bundle (`Fractal.jar`) ; ce bundle exporte des packages qui doivent être importés par les bundles contenant des composants Fractal.

FROGi utilise les mécanismes standard d'OSGi pour la gestion des activités de déploiement des bundles d'une application Fractal. Lors de l'installation des bundles, le framework OSGi résout de façon automatique les dépendances de code correspondant aux packages contenant les interfaces de service ainsi que l'API Fractal. Une fois ces dépendances résolues, le bundle peut être activé.

L'activation d'un bundle FROGi résulte dans l'instantiation et le démarrage d'une classe générique `FrogiBundleActivator` contenue dans chaque bundle FROGi. Cette classe est chargée de configurer l'environnement d'exécution Julia (notamment en précisant que le chargeur de classes à utiliser est celui du bundle) puis instancie une classe primaire (ie BootStrap) qui est chargée de créer les instances du ou des composants livrés dans le bundle. Il existe un singleton de cette classe par bundle, donc plusieurs sur une même plateforme OSGi.

*3.3. Gestion à l'exécution*

Cette section décrit les étapes de la gestion de l'activation des composants au moment de l'exécution.

*3.3.1. Publication de contrôleurs*

Une fois qu'une instance de composant FROGi (i.e., composant Fractal se trouvant à la racine d'un bundle) est créée, les interfaces de contrôle sont publiées dans le registre de services d'OSGi. L'enregistrement de ces interfaces permet à un autre bundle (son composite ou un bundle d'administration) de contrôler l'instance de composant.

Certains de ces contrôleurs peuvent également être enregistrés en adapter (*wrapper*) les interfaces Fractal à des interfaces standards du monde Java ou OSGi. C'est le cas par exemple du contrôleur *AttributeController* qui est adapté pour servir à la fois le service *javax.management.DynamicMBean* pour des agents JMX [FRE 04] et le service *org.osgi.service.cm.ManagedService* pour OSGi

*3.3.2. Liaison des instances*

Le courtage associé à l'approche orientée services est employé dans FROGi au moment de lier des instances de composants déployés dans des bundles différents. Un avantage d'utiliser la technique de courtage est qu'elle permet de réaliser des liaisons de façon flexible. Une liaison peut être réalisée, par exemple, par rapport à toute instance fournissant un service particulier (i.e *org.osgi.service.log.LogService*) caractérisés par une ensemble de propriétés (*langage=fr* ou *cron.pattern=\*\*\*3\*\*\** dans l'exemple des figures 2 et 3). Le courtage permet aussi de réaliser des liaisons 'rigides', dans ce cas une demande de service contient l'identificateur (*service.pid*) de l'instance vers laquelle la liaison doit être réalisée.

*3.3.3. Gestion du cycle de vie et des liaisons*

Le cycle de vie de l'instance d'un composant racine peut être gérée soit par son composite (qui est livré dans un autre bundle), soit par elle-même de façon autonome.

La gestion de cycle de vie par le composite impose premièrement que les interfaces serveurs de l'instance soient publiées comme des services dans le registre de services d'OSGi dès l'instanciation. Chaque service est identifié par l'attribut service.pid qui identifie l'instance du composant qui le fournit de façon unique et persistante. Le composite réalise les liaisons (*binding*) entre les instances qui le compose (au moyen des services *BindingController* des instances publiées dans le registre). Une fois les liaisons effectuées, le composite démarre les instances au moyen de leurs services *LifecycleController* publiés dans le registre.

L'alternative à cette politique est de considérer le bundle comme une unité autonome de gestion du cycle de vie de l'instance par rapport à son composite. Cette politique s'inspire de ServiceBinder (voir 4.2) qui assujettit le démarrage (méthode

*startFc()*) et l'arrêt (méthode *stopFc()*) de l'instance en fonction de la disponibilité des services correspondant aux interfaces clients, dans le registre de services OSGi. L'instance est démarrée dès que les dépendances de services obligatoires sont présentes dans le registre.

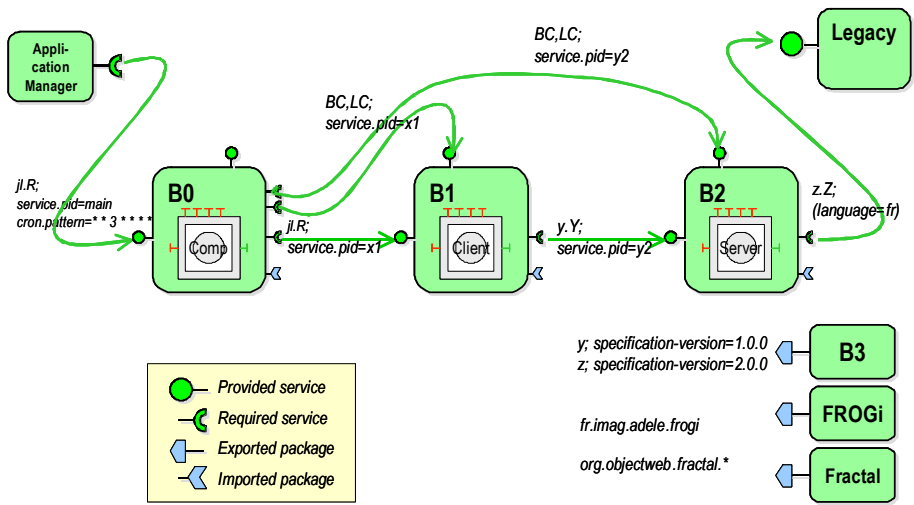

**Figure 2.** *Application Fractal mise sous forme de Bundles OSGi*

Cette dernière politique est utilisée pour connecter les composants à des bundles patrimoniaux qui sont dépourvus des contrôleurs de cycle de vie et de liaison.

### 3.3.4. Reconfiguration dynamique

Quelle soit la politique utilisée, une reconfiguration dynamique est nécessaire quand le framework notifie l'apparition d'un nouveau service ou le retrait d'un service utilisé par une instance. L'instance est premièrement arrêtée. La liaison ou la rupture est réalisée puis l'instance est redémarrée si les dépendances de services obligatoires sont présentes dans le registre. Dans le cas de la politique de gestion autonome, les services fournis sont systématiquement désenregistrés du registre OSGi au moment de l'arrêt et sont ré-enregistrés (toujours avec la même valeur de l'attribut *service.pid*) au redémarrage de l'instance.

### 3.2.5. Déclenchement de l'application

Une application Fractal est un composant/composite qui peut être lancée à partir d'une des interfaces fonctionnelles (e.g. *java.lang.Runnable*) par des outils comme le *org.objectweb.fractal.adl.Launcher*.

Dans OSGi, la notion d'application n'existe pas réellement : soit un bundle fournit un service qui est utilisé par un autre bundle (ie une commande shell, …) soit le bundle enregistre un objet auprès d'un service qui invoquera ultérieurement une méthode de cet objet (ie une servlet enregistrée auprès du *org.osgi.service.http.HttpService*), soit le bundle démarre une thread au moment de l'activation, soit le bundle repasse à l'état STOP après l'exécution normale de l'activateur.

Dans FROGi, un composant Fractal déployé est pris en charge par un ou plusieurs gestionnaires d'application résidant sur la passerelle en fonction de la « signature » du service. Ces gestionnaires déclenchent le composant en invoquant une méthode de service de l'interface client du composant. Par exemple, le temporisateur *CronService* invoque la méthode *run()* de services *java.lang.Runnable* ayant une propriété renseignée *cron.pattern*. De même, le Shell d'OSCAR invoque la méthode exécute sur le service *org.ungoverned.osgi.shell.Command* quand l'opérateur valide une commande à la console.

### *3.4. Extension de l'ADL Fractal*

FROGi propose des extensions à l'ADL Fractal pour tenir en compte les aspects liés au déploiement à l'intérieur de bundles OSGi.

Le sous-élément `<bundle>` des éléments `<component>` et `<description>` pilote le conditionnement de composant et de ses composants dans le bundle désigné par l'attribut `name`. L'attribut `version` spécifie la version de l'implémentation de l'ensemble. Il correspond à l'attribut `Bundle-Version` du manifeste du bundle. La récursion de conditionnement est interrompue par la rencontre d'un autre élément `<bundle>`.

L'attributs `bundle` de l'élément `<interface>` indique que l'interface doit être conditionnée dans un autre bundle dont le nom est spécifié par la valeur. Si la valeur de l'attribut bundle est la chaîne vide, l'interface n'est pas conditionnée par FROGi : elle est déjà disponible dans un autre bundle, généralement un bundle patrimonial. Par défaut, les interfaces de service sont conditionnées dans le même bundle que le composant où elles sont décrites. L'attribut `version` de l'élément `<interface>` indique la version de la spécification du package (ie contrat) de l'interface. La valeur par défaut de la version est 0.0.0.

Le sous-élément `<property>` de l'élément `<interface>` définit une des propriétés d'enregistrement du service correspondant l'interface serveur. Ces propriétés servent au courtage et aux gestionnaires d'application.

Le sous-élément `<binding>` des éléments `<component>` et `<description>` pilote à la fois la liaison « classique » des instances créées dans le même bundle, la liaison des instances créées dans des bundles séparées et la liaison d'une instance avec un service patrimonial OSGi. L'attribut `server` peut être substitué par l'attribut `filterserver` dont la valeur est une expression LDAP que doivent vérifier les propriétés du service à lier. Cet attribut n'est pas disponible pour les liaisons classiques (c-a-d intra-bundle). Remarquons que l'attribut `serverfilter="(service.pid=server.y2)"` est équivalent à `server="server.y2"`.

La figure 3 présente l'ADL Fractal nécessaire pour obtenir le conditionnement FROGi de la figure 2 pour l'application décrite dans la figure 1.

```xml
<definition name="HelloWorld">
  <bundle name="B0"/>
  <interface name="main" role="server" signature="java.lang.Runnable">
    <property name="cron.pattern" value="* * 3 * * * *"
                                 type="java.lang.String"/>
  </interface>
  <component name="client">
    <bundle name="B1"/>
    <interface name="x1" role="server" signature="java.lang.Runnable"/>
    <interface name="cy2" role="client" signature="y.Y"
        version="1.0.0" bundle="B3"/>
    <content class="ClientImpl"/> <!- but could be a composite -->
  </component>
  <component name="server">
    <bundle name="B2"/>
    <interface name="y2" role="server" signature="y.Y"
        version="1.0.0" bundle="B3"/>
    <interface name="cz3" role="client" signature="z.Z"
        cardinality="collection" contingency="optional"
        version="2.0.0" bundle="B3"/>
    <content class="ServerImpl"/> <!- but could be a composite -->
  </component>
  <binding client="this.x1" server="client.x1"/>
  <binding client="client.cy2" server="server.y2"/>
  <binding client="server.z3" server="this.cz3"/>
  <binding client="this.z3" serverfilter="(language=fr)"/>
</definition>
```

**Figure 3**. *Exemple d'ADL Fractal étendu par FROGi*

### 4. Travaux liés

Cette section présente divers travaux liés à l'utilisation d'OSGi comme une infrastructure pour déployer des composants ainsi qu'au conditionnement de composants Fractal.

#### *4.1. Beanome*

L'utilisation d'OSGi comme une infrastructure de déploiement de composants est explorée dans le modèle à composants Beanome [CER 02]. Dans Beanome, les bundles OSGi sont utilisés pour déployer des composants similaires à ceux de COM. De plus, le registre de services OSGi est utilisé pour publier les fabriques de composants au moment ou le bundle est activé. L'enregistrement de fabriques de composants en tant que services a pour avantage de permettre de localiser une fabrique non pas à partir d'un identifiant particulier comme dans le cas de COM, mais par rapport aux fonctionnalités fournies et requises par les composants instantiés par la fabrique. Beanome ne supporte cependant pas les changements dynamiques.

*4.2. Gravity*

Le projet Gravity [CER 04] explore la création d'applications capables de s'adapter de façon autonome par rapport à la disponibilité dynamique des composants qui les constituent. Gravity introduit un modèle à composants orienté services dans lequel le courtage est employé au moment de l'exécution pour lier les instances de composant mais aussi pour maintenir des compositions par rapport à l'arrivée et au départ de composants. Dans Gravity, une entité de l'environnement d'exécution, appelée le Service Binder, est chargée d'adapter les instances de composants et les compositions par rapport aux changements dynamiques. Gravity est implémenté comme une couche au dessus d'OSGi, et le Service Binder est déployé comme un bundle à l'intérieur de la plateforme de services. Un inconvénient de Gravity est qu'il utilise un modèle à composants particulier qui est cependant proche de Fractal.

*4.3. Paquetages Fractal*

Des discussions récentes dans la liste de diffusion de Fractal mentionnent la définition d'un moyen de conditionner des composants Fractal. Ces concepts, qui sont actuellement dans un stade de proposition, ne considèrent cependant pas l'existence d'une infrastructure permettant de réaliser des activités de déploiement continues.

## 5. Conclusion

Cet article a présenté FROGi, une proposition visant à introduire des caractéristiques de la plateforme de services OSGi dans le modèle à composants Fractal. Dans FROGi, une application Fractal est conditionnée sous la forme d'un ou plusieurs bundles OSGi ; ce qui permet de réaliser leur livraison et leur déploiement de façon individuelle et continue. De plus, la liaison entre les instances de composants peut être réalisée soit à travers la technique de connexion 'standard' de Fractal, soit à partir de la publication des interfaces fonctionnelles dans le registre de services et l'utilisation de la technique de courtage propre à OSGi. FROGi propose en plus des extensions à l'ADL Fractal pour permettre de prendre en compte les aspects de conditionnement. L'expérimentation réalisée autour de FROGi à permis de comprendre les limitations de l'utilisation de Julia dans un contexte de classloaders multiples tel que celui d'OSGi. Ces limitations ont été exposées aux développeurs de Julia qui, en réponse, ont proposé des solutions. Finalement, il faut souligner que FROGi ainsi que différents travaux énoncés dans la section 4 montrent qu'OSGi est une plateforme idéale pour réaliser le déploiement de composants. FROGi est disponible à l'adresse http://www-adele.imag.fr/frogi

Cependant divers points qui n'ont pu être traités dans les travaux réalisés jusqu'à maintenant et font l'objet de nos recherches actuelles.

**Mécanisme de création d'instances multiples :** Fractal supporte la création d'un nombre variable d'instances de composant. Le travail présenté ici est focalisé sur une approche basée sur l'utilisation de singletons. Un moyen de résoudre ceci, tout en restant compatibles avec l'environnement OSGi, est de publier des fabriques de composants au moyen de services (comme dans l'approche de la section 4.1)

**Support des changements dynamiques :** Un besoin qui s'impose au moment d'utiliser la plateforme OSGi est le support des changements dynamiques. Bien que Fractal soit orienté vers le support de la disponibilité dynamique, ceci doit être géré au niveau de la programmation. Une approche comme celle suivie dans le Service Binder (voir section 4.2) pourrait être utilisée pour résoudre ceci.

**Génération automatique à partir de l'ADL :** Un travail qui reste à réaliser est celui de la génération automatique de bundles à partir de l'ADL étendu présenté dans la section 3.4.

## 7. Bibliographie